# Machine-learning-accelerated discovery of synthesizable high-temperature altermagnets with giant spin splitting

Yi-Fei Jiang[1,2], Jia-Xuan Guo[1,2], Zhen Zhang[3], Xin-Wei Yi[4]*, Jing-Yang You[2]*

[1] School of Physics, Beihang University, Beijing 100191, China

[2] Peng Huanwu Collaborative Center for Research and Education, Beihang University, Beijing 100191, China

[3] State Key Laboratory of Quantum Optics and Quantum Optics Devices, Institute of Theoretical Physics, Shanxi University, Taiyuan 030006, China

[4] Institute of Theoretical Physics, Chinese Academy of Sciences, Beijing 100190, China

*Correspondence and requests for materials should be addressed to Xin-Wei Yi (yixinwei@itp.ac.cn) and Jing-Yang You (phyjyy@buaa.edu.cn)

## Abstract

Altermagnets offer a route to spin-polarized electronic states without macroscopic magnetization, because their compensated magnetic order can generate momentum-dependent spin splitting through crystal-symmetry-controlled exchange fields. However, experimentally viable altermagnets that combine large spin splitting, thermodynamic stability and high magnetic ordering temperatures remain scarce. Here, we develop a machine-learning-accelerated high-throughput framework to explore the tetragonal $AB_2C_2D$ materials space. Screening 8640 structural and chemical variants identifies 1347 compensated antiferromagnetic candidates satisfying the symmetry requirements for altermagnetism. An interpretable XGBoost model trained on first-principles spin-splitting data then isolates 34 low-hull-energy candidates, four of which have been previously reported, with giant non-relativistic spin splittings exceeding 1.5 eV near the Fermi level. Detailed first-principles calculations of the representative $RbMn_2Te_2O$ confirm a maximum spin splitting of ~1.88 eV with dynamical stability and an estimated Néel temperature of ~390 K. The giant splitting originates from symmetry-locked Mn-sublattice exchange fields amplified by directional Mn-*d*/Te-*p* hybridization. Furthermore, we uncover a profound soft-mode-driven structural transition associated with an interlayer dimensionality crossover in $SrMn_2Te_2O$, yet the unfolded electronic structure demonstrates that the altermagnetic spin splitting remains robust after lattice reconstruction. Hydrostatic pressure provides an additional tuning route, producing non-monotonic modulation of the spin-split Fermi surface governed by local coordination and orbital hybridization. These results establish tetragonal $AB_2C_2D$ compounds as a tunable materials platform for stray-field-free spintronic devices and provide a general data-driven strategy for discovering robust giant-splitting altermagnets.

## Introduction

Spintronics seeks to use the electron spin, in addition to charge, for information storage, transport, and manipulation[1,2]. Ferromagnets naturally provide exchange-split bands and spin-polarized carriers, but their finite magnetization generates stray magnetic fields that limit device miniaturization and high-density integration. Antiferromagnets avoid this problem because their magnetic moments compensate macroscopically and their spin dynamics can be fast[3,4]. In conventional collinear antiferromagnets, however, combined space-time symmetries often protect spin degeneracy throughout the Brillouin zone, limiting their ability to generate large spin-polarized transport responses[5].

Altermagnets have recently emerged as a distinct form of compensated magnetic order that overcomes this limitation[6,7]. In an altermagnet, opposite-spin sublattices are connected not by simple inversion or translation combined with time reversal, but by crystal rotations or mirrors that imprint the magnetic order onto momentum space[6–8]. The result is a spin-split band structure with zero net magnetization. Because the splitting is produced primarily by non-relativistic exchange rather than by spin-orbit coupling, it can reach energy scales far exceeding those typical of relativistic spin splitting. This combination of compensated magnetism, large spin polarization and symmetry-controlled momentum texture makes altermagnets promising for stray-field-free, high-speed, and low-power spintronic devices[5,9–12].

A central obstacle is that only a limited number of materials are currently known to meet the simultaneous requirements for device-relevant altermagnetism: giant spin splitting near the Fermi level, high magnetic ordering temperatures, thermodynamic and dynamical stability, and chemical synthesizability[11,13]. These requirements are strongly coupled. The magnitude of altermagnetic spin splitting depends on the magnetic sublattice, orbital character, local crystal filed and exchange pathways, and the same ingredients also control phase stability and magnetic ordering[6,14]. Consequently, conventional trial-and-error searches are inefficient, especially in multicomponent chemical spaces where small changes in composition can strongly reshape the lattice geometry and orbital hybridization.

The tetragonal $AB_2C_2D$ family, represented by prototypes such as $KV_2Se_2O$ and $RbV_2Te_2O$[15,16], provides an ideal platform for addressing this challenge. In this structure, the B-site transition metals form the magnetic network, whereas the A-, C-, and D-site elements tune the interlayer spacing, charge balance, ligand environments and orbital overlap. This separation of roles allows magnetic exchange and structural stability to be adjusted through chemically distinct degrees of freedom. Importantly, when the tetragonal lattice is paired with appropriate compensated antiferromagnetic (AFM) configurations, the crystallographic and magnetic symmetries satisfy the conditions required for altermagnetism[7].

Here we combine high-throughput first-principles calculations with interpretable machine-learning (ML) to map the altermagnetic landscape of tetragonal $AB_2C_2D$ compounds[17,18]. We screen 8640 systematically substituted compounds, evaluate competing magnetic configurations and identify 1347 compensated AFM candidates. A descriptor-based XGBoost model[19] is then trained to predict the maximum altermagnetic spin splitting, enabling rapid identification of low-hull-energy compounds with giant non-relativistic spin splitting. Beyond materials discovery, we use representative compounds to uncover the microscopic origin, structural robustness and external tunability of the giant splitting. $RbMn_2Te_2O$ demonstrates a stable room-temperature altermagnet in which symmetry-locked Mn exchange fields are amplified by directional Mn-*d*/Te-*p* hybridization. $SrMn_2Te_2O$ reveals that soft-mode structural reconstruction can drive a dimensionality crossover while preserving the essential altermagnetic band topology. Pressure calculations further show that the spin-split Fermi surface can be tuned non-monotonically by controlling local bonding and orbital hybridization. Together, these results establish $AB_2C_2D$ compounds as a broad and chemically tunable family of robust giant altermagnets.

## Results

### ML-accelerated high-throughput discovery of $AB_2C_2D$ altermagnets

To explore altermagnetism across a large but chemically structured materials space, we designed a hierarchical high-throughput and ML workflow for tetragonal $AB_2C_2D$ compounds derived from the $KV_2Se_2O$ prototype (Fig. 1). The workflow consists of three stages: (i) first-principles screening of structural and magnetic ground states, (ii) ML-accelerated prediction of maximum altermagnetic spin splitting, and (iii) final selection of candidates that combine thermodynamic stability with giant spin splitting.

The $AB_2C_2D$ framework is particularly useful because its structural and magnetic degrees of freedom can be tuned semi-independency. The B-site transition metals carry the local moments and determine the dominant exchange interactions. By contrast, the nominally non-magnetic A, C, and D sites control lattice constants, coordination polyhedra, ligand fields and orbital overlaps. This site-selective tunability makes it possible to search for compounds in which the symmetry of a compensated AFM state generates altermagnetic spin splitting, while the surrounding chemical environment optimizes both stability and the magnitude of the splitting.

In the first stage (Fig. 1a), we constructed 8640 $AB_2C_2D$ compounds through systematic elemental substitution. Each structure was relaxed using first-principles calculations, and its nonmagnetic, ferromagnetic, and AFM configurations were compared by total energy. We then combined energetic selection with magnetic-symmetry analysis to identify compounds that form fully compensated AFM states and satisfy the crystallographic requirements for altermagnetism[6,7]. This procedure reduced the full chemical space to 1347 candidate altermagnets.

Directly calculating spin-resolved band structures and maximum spin splittings for all candidates would be computationally expensive. We therefore introduced a ML model in the second stage to accelerate the search. An XGBoost regression model was trained to map compositional, structural, and magnetic descriptors onto the maximum altermagnetic spin splitting ($\Delta E_X$) (Fig. 1b). The model was trained on 80% of the candidate set with first-principles-calculated $\Delta E_X$ values and evaluated on a strictly held-out 20% test set. The predicted and DFT-calculated spin splittings are in close agreement (Fig. 1c), showing that the model captures the main chemical and structural trends governing the splitting. Under 10-fold nested cross-validation[20], the model achieves a remarkable coefficient of determination ($R^2 = 0.83$), a mean absolute error (MAE) of 0.17 eV and a root-mean-square error (RMSE) of 0.25 eV.

We used SHAP (SHapley Additive exPlanations)[21,22] analysis to interpret the model and connect its predictions to physical mechanisms (inset of Fig. 1c). The most important descriptors are the periodic-table row of the B-site element, the minimum relative distance at Sites C and D, the Allen electronegativity[23] and covalent radius of the B-site element. The minimum relative distance is defined as the shortest local interatomic distance around a given site normalized by the sum of the estimated ionic radii, $\min_j[R_{ij}/(r_i + r_j)]$, and thus measures the degree of local bond compression. The leading importance of the B-site row confirms that the orbital radial extent, crystal-field response and exchange scale of magnetic *d* shell are central to giant altermagnetic splitting. The strong contribution from the C/D-site relative distances indicates that the effect is highly sensitive to ligand coordination, because small changes in metal-ligand bond lengths and bond angles alter the anisotropic *d*-*p*-*d* hopping channels. Electronegativity and covalent radius further encode charge transfer and chemical pressure. Overall, the model shows that large $\Delta E_X$ in $AB_2C_2D$ compounds is not controlled by a single elemental property, but by the cooperative effect of magnetic-site chemistry, ligand-field distortion and directional orbital hybridization.

The prediction accuracy also depends strongly on thermodynamic stability. When the energy above the convex hull, $E_{hull}$, is below 0.1 eV/atom, the RMSE decrease to ~0.2 eV. In the high-performance region with $\Delta E_X > 1.5$ eV, the error is further reduced to ~0.1 eV. This trend has a clear physical origin. Low-$E_{hull}$ structures occupy chemically reasonable minima on the potential-energy surface, with stable coordination environments, moderate internal strain and smooth variations in orbital hybridization under substitution. These conditions make the descriptor-property relationship more continuous and therefore more learnable. By contrast, high-$E_{hull}$ structures often contain anomalously short bonds, severe distortions or non-equilibrium coordination environments. Such structures can produce very large but discontinuous spin splittings, occasionally exceeding 3 eV, because extreme orbital overlaps push the electronic structure outside the distribution represented by the stable compounds. The improved model reliability in the low-$E_{hull}$ regime is therefore advantageous: it is highest precisely in the part of chemical space most relevant for synthesis.

In the final stage, we applied the validated model across the candidate set and filtered the results using both thermodynamic stability and spin-splitting magnitude. Figure 1d plots $E_{hull}$ against the predicted maximum spin splitting. Among the 1347 symmetry-qualified altermagnetic candidates, 51 compounds fall within the low-hull-energy window of $E_{hull} \leqslant 0.1$ eV/atom, and 34 of these are predicted to exhibit giant spin splitting exceeding 1.5 eV. Notably, this subset includes four materials that have been previously reported or experimentally observed, namely $CsV_2Te_2O$, $KV_2Se_2O$, $CsV_2Se_2O$, and $RbV_2Se_2O$[15,24–26], the full list of compounds is provided in Supplementary Table S1. These compounds form a compact set of experimentally relevant targets because they combine energetic accessibility with large exchange-driven spin polarization.

To benchmark the model against known altermagnetic compounds, we compared its predictions with spin splittings extracted from previously reported $AB_2C_2D$ materials (Fig.

1e)[15,27,28]. The agreement with literature band-structure results supports the transferability of the descriptor-based model within this structural family and validates its use for identifying unexplored giant-splitting altermagnets.

**$RbMn_2Te_2O$: a dynamically stable altermagnet with giant spin splitting**

We selected $RbMn_2Te_2O$, one of the most stable compounds among the 34 high-performance candidates, for detailed first-principles characterization. The relaxed structure retains a layered tetragonal structure (Fig. 2a). Mn atoms, coordinated by O and Te-derived ligand environments, form the 2D magnetic sublattice, whereas Rb acts primarily as an interlayer spacer that stabilizes the lattice and tunes the crystal field. The two Mn sublattices carry antiparallel local moments of equal magnitude, producing a fully compensated AFM state with zero macroscopic magnetization. Despite this compensation, unlike conventional symmetry-protected antiferromagnets, the opposite-spin Mn sublattices in $RbMn_2Te_2O$ is not a conventional spin-degenerate antiferromagnet. The opposite-spin Mn sublattice are not related by inversion or translation combined time-reversal. Instead, they are connected by $[C_2 \, || \, C_{4z}]$, which couples spin reversal to a real-space rotation of the local orbital environment[6]. This symmetry relation removes the global spin degeneracy away from special momenta and enforces a momentum-dependent exchange splitting. Because the mechanism is non-relativistic, the splitting is controlled by exchange and orbital hopping rather than by the much smaller spin-orbit energy scale.

The spin-resolved band structure calculated without spin-orbit coupling confirms this altermagnetic electronic structure (Fig. 2b). Along the Γ-X-M-Y-Γ path, the two opposite spin channels are strongly split while the total magnetization remains zero. The maximum separation reaches ~1.88 eV near the X/Y valleys. This energy scale is far larger than typical relativistic spin splittings[29] and is consistent with an exchange-driven origin.

The projected density of states (PDOS) reveals the microscopic source of the large splitting. The electronic states near the Fermi level are dominated by Mn-*d* orbitals hybridized with Te-*p* ligand states. The Mn sublattices provide the exchange field, but the ligand network amplifies the altermagnetic response by making the hopping strongly direction dependent. In a tetragonal environment, Mn $d_{xz}$ and $d_{yz}$ orbitals couple differently along the $k_x$ and $k_y$ directions through Te-*p*-mediated pathways. When the two spin sublattices are related by a rotation that interchanges these orbital orientations, the orbital anisotropy becomes locked to spin reversal. The result is a large spin splitting that changes sign or magnitude with momentum, rather than a uniform ferromagnetic exchange shift.

This mechanism is visible in the constant-energy maps and fat-band projections. The spin-resolved constant-energy contours in the $k_x$-$k_y$ plane over an energy window of -0.5 to 0.5 eV (Fig. 2c) show alternating spin character along the $k_x$ and $k_y$ directions, consistent with a *d*-wave-like altermagnetic spin texture. Sublattice- and spin-resolved fat bands (Figs. 2d, e) show that the bands responsible for the maximum splitting at X are dominated by the $d_{x^2-y^2}$ orbitals from the two antiparallel Mn sublattices, with minor Mn-$d_{z^2}$ weight (Supplementary Fig. S1). The adjacent spin-split band (Fig. 2e) mainly originates from the spin up Mn1-$d_{yz}$ and spin-down Mn2-$d_{xz}$ orbitals. These orbital assignments indicate that the giant splitting is generated by the combined action of sublattice exchange and anisotropic orbital hopping, not merely by the size of the local Mn moment.

The symmetry interpretation is as follows. The operation connecting the two Mn sublattices reverses the spin and rotate the local orbital basis. Under the $C_{4z}$, the $d_{xz}$ and $d_{yz}$ transform into one another up to a phase. Thus, an orbital texture dominated by $d_{yz}$ on the one spin sublattice is mapped to a $d_{xz}$ texture on the opposite-spin sublattice. In reciprocal space this produces unequal spin-channel dispersions along $k_x$ and $k_y$, while preserving zero net magnetization in the unit cell. The fat-band projections therefore provide a direct microscopic visualization of the symmetry-locked spin-orbital texture that defines altermagnetism in $RbMn_2Te_2O$.

$RbMn_2Te_2O$ also satisfies the stability requirements for practical altermagnetic materials. Phonon calculations show no imaginary frequencies, confirming dynamical stability. Magnetic exchange parameters extracted from first-principles calculations and mapped onto an effective spin Hamiltonian yield, through atomistic Monte Carlo simulations[30,31], an estimated Néel temperature ($T_N$) of ~390 K (Supplementary Fig. S2). The coexistence of dynamical stability, above-room-temperature AFM order and a 1.88 eV non-relativistic spin splitting makes $RbMn_2Te_2O$ a strong candidate for experimental realization and spintronic applications.

**$SrMn_2Te_2O$: soft-mode-driven structural transition and robustness of altermagnetic splitting**

The high-throughput search also identifies low-energy $AB_2C_2D$ whose pristine high-symmetry phases show lattice softening. $SrMn_2Te_2O$ is a representative case for studying how structural instability interacts with altermagnetic electronic topology. Replacing interlayer Rb with Sr changes the ionic charge, ionic radius and interlayer electrostatics. These changes alter both the band filling and the balance between quasi-two-dimensional layering and interlayer bonding, making the pristine tetragonal phase susceptible to a structural distortion.

Phonon calculations for the high-symmetry $SrMn_2Te_2O$ phase show pronounced imaginary frequencies at the Brillouin-zone boundary near X and Y when an electronic smearing of $\sigma$ = 0.05 eV is used (Fig. 3a). This mode signifies that the pristine phase is dynamically unstable at 0 K. In metallic or semimetallic systems, such zone-boundary soft modes are often associated with Fermi surface nesting or strong momentum-selective electron-phonon coupling (EPC)[32–34]. We therefore calculated the bare electronic susceptibility (nesting function) and the momentum-resolved EPC. Neither quantity shows a pronounced anomaly at **q** = (0.5, 0, 0), suggesting that the instability is not a conventional nesting-driven charge density wave (CDW). Instead, the soft mode is primarily structural. Sr substitution enhances the tendency toward interlayer A-O interactions and creates a steric and electrostatic mismatch within the Te-O coordination framework. The lattice lowers its energy by buckling the initially planar oxygen network and allowing the A-site cations to move along the *c* direction toward the oxygen cage. This motion converts part of the weakly coupled layered structure into a more three-dimensionally connected bonding network (Supplementary Fig. S4). The soft mode can therefore be interpreted as a structural dimensionality crossover from a quasi-two-dimensional pristine lattice to a more strongly bonded three-dimensional distorted phase.

Although the instability is not driven by a conventional nesting mechanism, it remains sensitive to electronic occupation near the Fermi level. Increasing the electronic smearing from 0.01 eV to 0.4 eV progressively hardens the imaginary modes and eventually removes them (Fig. 3a). Electronic smearing is not equivalent to lattice temperature, but this trend shows that low-energy electronic states participate in the energy balance. The distortion likely produces an electronic energy gain by lifting unfavorable degeneracies or reducing the DOS near the Fermi level, akin to a band-Jahn-Teller effect[35], while the primary displacement pattern is set by lattice geometry and interlayer bonding.

To identify the low-temperature ground state, we froze in the unstable X/Y soft-mode eigenvectors in a corresponding 2×2×1 supercell and followed the total energy as a function of the distortion amplitude (Fig. 3b). The energy decreases strongly along the soft-mode coordinate and the structure relaxes into a low-symmetry phase. This confirms that the imaginary phonon is not a numerical artifact but represents a genuine structural reconstruction. The relaxed distorted supercell has no imaginary frequencies (Fig. 3d), demonstrating that it is dynamically stable.

We further evaluated the temperature-dependent Helmholtz free-energy difference between the distorted and pristine phases, $\Delta F(\mathrm{T}) = F_{\mathrm{dist}}(\mathrm{T}) - F_{\mathrm{pri}}(\mathrm{T})$ (Fig. 3c). The distorted phases is favored at low temperature, whereas the pristine phase becomes thermodynamically competitive at elevated temperature due to vibrational entropy and electronic thermal effects. The calculated crossing at ~1390 K indicates an entropy-stabilized high-symmetry phase at high temperature, although the precise transition temperature should be interpreted within the approximations of the harmonic free-energy treatment.

The key question is whether the altermagnetic spin splitting survives this lattice reconstruction. To answer this, we calculated the spin-resolved band structure of the distorted 2×2×1 supercell and unfolded it onto the Brillouin zone of the pristine high-symmetry cell (Fig. 3e and Supplementary Fig. 3)[36,37]. The unfolded spectra reveal that the spin-split Mn-*d* bands remain strongly asymmetric in momentum spate after distortion. Thus, the structural transition lowers the elastic and bonding energy but does not destroy the magnetic-symmetry mechanism that produces altermagnetism. The essential reason is that the distortion preserves the core relation between opposite-spin Mn sublattices and their anisotropic orbital environments, even though the translational symmetry is reduced.

The soft-mode mechanism also reveals an effective route for external control. Since the distortion is governed by the competition between interlayer A-O bonding and the quasi-two-dimensional pristine framework, the instability is highly sensitive to lattice anisotropy, particularly the $c/a$ ratio. As shown in Fig. 3f, the energy difference $\Delta E = E_{\mathrm{dist}} - E_{\mathrm{pri}}$ increases toward zero under $c$-axis tensile strain, indicating that enlarged interlayer spacing weakens the driving force for oxygen buckling. This suppression is more efficient when the in-plane lattice constants are allowed to relax, demonstrating that the instability is controlled by the coupled response of interlayer separation and in-plane bonding rather than by the $c$-axis length alone. By contrast, isotropic tensile strain does not suppress the distortion, further highlighting the central role of $c/a$. Consistently, electron doping expands the lattice anisotropically, with a larger relative increase along the $c$ axis than in the $ab$ plane, thereby increasing $c/a$ and suppressing the structural transition (Supplementary Fig. S4). These results identify lattice anisotropy as a practical control parameter for tuning the soft-mode dimensionality crossover through strain, pressure, chemical substitution, intercalation, or electrostatic doping, while maintaining the robust altermagnetic electronic structure.

**Global candidate landscape, tunable giant spin splitting and spintronic outlook**

To summarize the screening results in a form useful for experimental prioritization, we constructed a stability–magnetism map for the candidate $AB_2C_2D$ altermagnets (Fig. 4a). The map combines the $E_{hull}$, local magnetic moment and estimated $T_N$. Within the stringent stability window of $E_{hull} \leq 0.1$ eV/atom, a substantial group of compounds exhibits both robust local magnetic moments and high estimated $T_N$. This region identifies candidates that are not only electronically attractive, but also likely to be accessible for synthesis and finite-temperature measurements.

Device applications require not only large intrinsic spin splitting but also external control. We therefore calculated the maximum spin splitting under hydrostatic pressure for $SrMn_2Te_2O$ and $RbMn_2Te_2O$ (Fig. 4b). Both systems exhibits non-monotonic pressure dependence, proving that the giant altermagnetic splitting is a tunable band-structure property rather than a fixed material constant. In $SrMn_2Te_2O$, the maximum spin splitting remains above 1.5 eV and increases beyond 2.0 eV at 50 GPa. In $RbMn_2Te_2O$ , the splitting decreases toward a minimum near 20 GPa and increases again under stronger compression. The microscopic origin of this pressure response is the sensitivity of anisotropic exchange splitting to local bonding.

Sublattice- and orbital-resolved fat bands for $RbMn_2Te_2O$ at 0 and 40 GPa (Fig. 4c) reveal that pressure shifts the relative energies, bandwidths and orbital weights of Mn-derived states. Compression modifies Mn-ligand distances and bond angles, which changes the crystal-field splitting and the directional Mn-*d*/ligand-*p* hybridization pathways. Because the altermagnetic splitting is generated by the momentum-dependent difference between these orbital hopping channels on opposite spin sublattices, pressure can either enhance or reduce the splitting depending on how it redistributes the relevant orbital character near the Fermi level. The non-monotonic trend therefore reflects competition among bandwidth broadening, crystal-field reordering and exchange-driven spin polarization.

Although the hydrostatic pressures used here map an extreme tuning range, the same physical mechanism can be accessed by experimentally realistic perturbations. Epitaxial strain can selectively modify in-plane and out-of-plane bond geometries; chemical substitution can apply chemical pressure and tune charge transfer; and heterostructure interfaces can reshape coordination, screening and orbital alignment. These routes provide practical handles for engineering altermagnetic spin textures in thin films and devices.

The combination of zero net magnetization and giant spin splitting suggests a natural device concept: an altermagnetic tunnel junction (AM-MTJ) based on $AB_2C_2D$ electrodes separated by an insulating barrier (Fig. 4d). Conventional ferromagnetic MTJs provide large tunneling magnetoresistance but suffer from stray-field-induced crosstalk[38]. Conventional compensated antiferromagnets remove stray fields, but their spin-degenerate bands make electrical readout challenging. $AB_2C_2D$ altermagnets bridge these limits. Their compensated magnetic order suppresses stray fields, while their momentum-dependent spin splitting provides a spin-selective electronic structure. Reversing the Néel vector in one electrode reverses the spin texture in momentum space, changing the spin and momentum matching across the barrier. In a suitably oriented junction, this can generate a large change in tunneling conductance through momentum-resolved spin filtering. The $AB_2C_2D$ family therefore offers a materials basis for high-density, non-volatile and potentially ultrafast altermagnetic memory devices.

## Discussion

We have developed a ML-accelerated high-throughput framework for discovering giant-splitting altermagnets in tetragonal $AB_2C_2D$ compounds. Starting from 8640 structural and chemical variants, the workflow identifies 1347 compensated antiferromagnetic candidates satisfying the symmetry requirements for altermagnetism and isolates 34 low-hull-energy compounds predicted to host spin splittings above 1.5 eV. The interpretable ML analysis shows that the magnitude of the splitting is controlled by the cooperative effect of magnetic-site *d*-shell chemistry, local coordination geometry, charge transfer and directional *p*-*d* hybridization.

First-principles characterization of representative materials reveals the underlying physical mechanisms. $RbMn_2Te_2O$ is dynamically stable, has an estimated $T_N$ of about 390 K and exhibits a maximum non-relativistic spin splitting of about 1.88 eV. Its giant splitting arises from the symmetry-locked relationship between antiparallel Mn sublattices and anisotropic Mn-*d* orbital textures, with Te-*p*-mediated hopping amplifying the momentum-dependent exchange splitting. $SrMn_2Te_2O$ shows that soft-mode lattice reconstruction can lower the structural energy through an interlayer dimensionality crossover while preserving the essential altermagnetic band topology. Pressure calculations further demonstrate that the spin-split Fermi surface can be tuned non-monotonically through local coordination and orbital-hybridization control.

These findings move beyond a catalogue of candidate materials. They establish design principles for robust altermagnets: choose magnetic sublattices with strong exchange and anisotropic *d* orbitals; use ligand and spacer chemistry to tune directional hybridization and stability; and exploit strain, pressure or chemical substitution to control the spin-split Fermi surface. The $AB_2C_2D$ family therefore provides both a rich materials library for experimental synthesis and a tunable platform for realizing stray-field-free spintronic devices based on giant exchange-driven spin splitting.

## Methods

### Construction of the $AB_2C_2D$ compositional space

We constructed a large candidate library based on the tetragonal $AB_2C_2D$ prototype structure. As illustrated in Fig. 5, the primitive cell contains four chemically distinct crystallographic sites, allowing targeted elemental substitution at each sublattice. The interlayer A site was substituted by 10 alkaline and alkaline-earth metals (Li, Be, Na, Mg, K, Ca, Rb, Sr,

Cs, Ba); the magnetic B site by 27 transition metals ranging from periods 4 to 6 (Sc through Au); the C site by 8 pnictogens and chalcogens (O, S, Se, Te, P, As, Sb, Bi); and the D site by 4 chalcogens (O, S, Se, Te). Exhaustive site-specific permutation generated an initial compositional space comprising 8640 $AB_2C_2D$ candidate structures.

**High-throughput first-principles calculations**

All first-principles calculations were performed using density functional theory (DFT) as implemented in the Vienna Ab initio Simulation Package (VASP)[39]. The electron-ion interactions were described by the projector augmented wave (PAW) method[40], and the exchange-correlation interactions were treated within the generalized gradient approximation (GGA) using the Perdew–Burke–Ernzerhof (PBE) functional[41]. Unless otherwise specified, SOC was excluded in the high-throughput screening, allowing us to isolate the nonrelativistic anisotropic exchange splitting characteristic of altermagnetism.

For the high-throughput screening workflow, the plane-wave energy cutoff was set to 520 eV. Brillouin-zone integration was performed using Γ-centered k-point meshes[42] with a typical density of KSPACING = 0.2 $Å^{-1}$. The energy convergence criterion for the electronic self-consistence loop was set to $10^{-5}$ eV. Full structural relaxations (including both lattice parameters and atomic positions) were performed until the residual forces on each atom dropped below 0.01 eV/Å. To evaluate the thermodynamic stability ($E_{hull}$) in a manner compatible with the Materials Project database[43,44], the GGA+$U$ method[45] was adopted during the high-throughput screening stage. Hubbard corrections were applied to the d orbitals of the B-site transition-metal elements using species-specific $U_{eff}$ values calibrated by the Materials Project[46,47]. This protocol enables a consistent comparison of the thermodynamic stability of chemically diverse $AB_2C_2D$ compounds.

For each composition, several competing magnetic configurations, including nonmagnetic, ferromagnetic, and antiferromagnetic states, were fully relaxed and compared in total energy. Candidate altermagnets were first selected by requiring a fully compensated antiferromagnetic ground state with nearly zero net magnetic moment. The selected compounds were then subjected to magnetic symmetry analysis to confirm that the compensated spin sublattices satisfy the symmetry requirements for altermagnetism, namely the absence of a global symmetry that enforces spin degeneracy throughout the Brillouin zone and the presence of crystallographic rotations connecting opposite-spin sublattices.

After identifying 1347 symmetry-allowed compensated antiferromagnetic candidates, spin-resolved band structures were calculated without Hubbard $U$ corrections. This parameter-free PBE treatment provides a uniform reference for comparing nonrelativistic altermagnetic spin splittings across a chemically broad materials space, avoiding artificial variations introduced by empirical $U$ values. The validity of this choice was benchmarked using the prototypical compound $KV_2Se_2O$, for which the PBE-calculated Fermi surface reproduces the main features observed in experimental angle-resolved photoemission spectroscopy (ARPES) [Supplementary Fig. S5][15], including the distribution of V-3$d$ states at the Fermi level.

Phonon spectra were calculated using the finite-displacement method[48] as implemented in Phonopy[49,50], with interatomic force constants obtained from VASP calculations. The force calculations employed a plane-wave energy cutoff of 400 eV, a dense k-mesh of 2 × 2 × 2, and 3 × 3 × 2 supercells. The absence of imaginary phonon modes was used as the criterion for dynamical stability.

**Machine-learning workflow and feature engineering**

To accelerate the search for giant-spin-splitting altermagnets, we developed a supervised ML regression model to predict the maximum nonrelativistic altermagneric spin splitting near the Fermi level ($\Delta E_X$). The target $\Delta E_X$ was obtained from spin-resolved PBE band structures and defined as the maximum energy separation between paired opposite-spin bands at the X point, within an energy window of $E_F \pm 0.6$ eV.

The full dataset was divided into a training/validation subset containing ~80% of the DFT-labelled candidates and an independent hold-out subset containing the remaining 20%. The hold-out subset was not used during model training, feature selection, or hyperparameter optimization.

The input feature space was constructed from structural and compositional descriptors generated via matminer[51]. These descriptors were further augmented by physically motivated quantities, including B-site local magnetic moments, OctaDist[52]-derived coordination-distortion parameters, and site-dependent electronegativities. To reduce redundancy and improve model transferability, highly correlated descriptors were removed, and recursive feature elimination with cross-validation was applied to identify an optimal subset of predictive features.

An XGBoost[19] regression model was used to capture the nonlinear relationship between chemistry, local structure, magnetic moments, and altermagnetic spin splitting. The ML workflow was implemented using scikit-learn library[53], with hyperparameter optimization was performed using Optuna[54]. Model performance was evaluated through 10-fold nested cross-validation, in which the inner loop was used for hyperparameter tuning and the outer loop for was used for unbiased out-of-sample evaluation. The predictive accuracy was quantified using the $R^2$, MAE, and RMSE. After validation, the optimized model was applied to the independent hold-out set to predict the spin-splitting magnitude of previously unlabeled candidates and to identify thermodynamically stable compounds with giant altermagnetic spin splitting

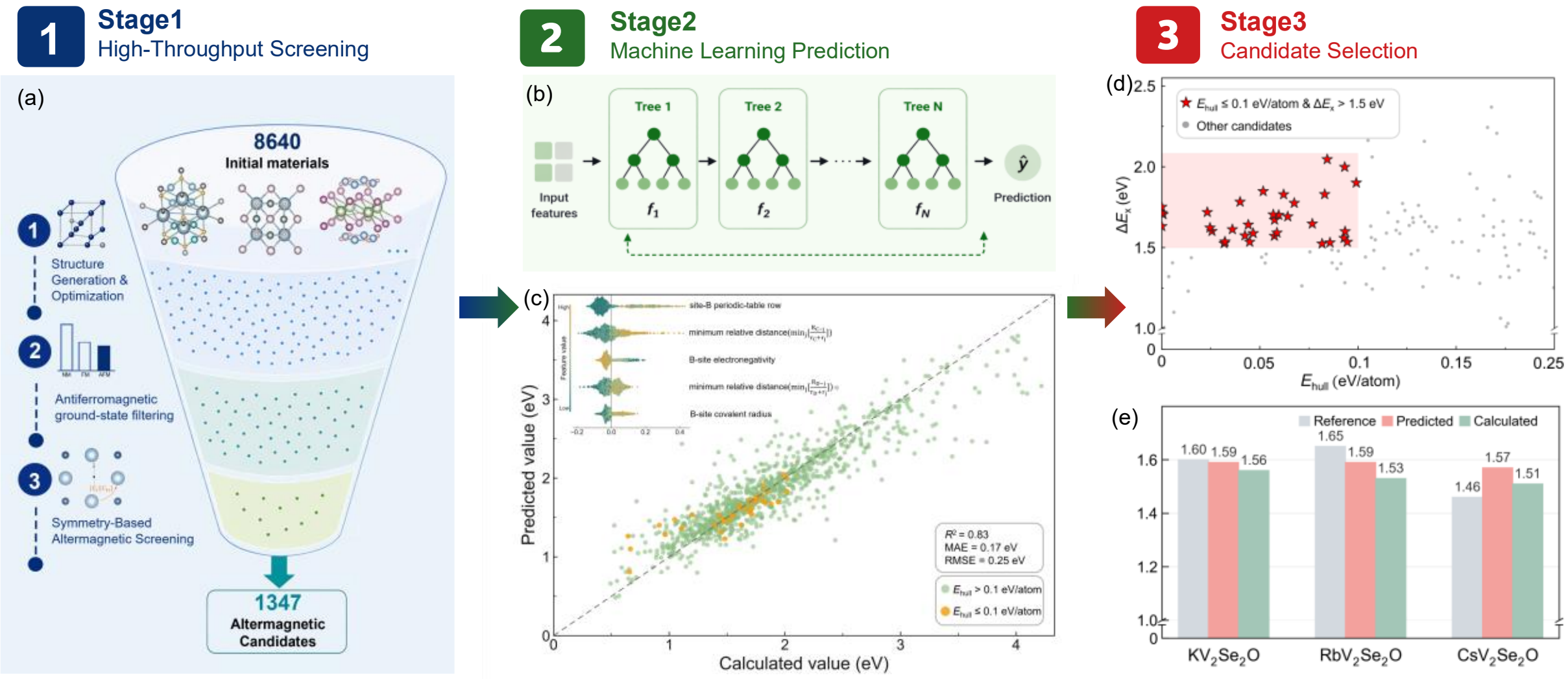

**Fig. 1 | Machine-learning-accelerated high-throughput discovery of giant-spin-splitting $AB_2C_2D$ altermagnets. a,** Schematic workflow of the high-throughput screening procedure. Exhaustive site-specific elemental substitution of the tetragonal $AB_2C_2D$ prototype generates an initial chemical space of 8640 candidate structures. Successive filters based on first-principles structural relaxation, magnetic ground-state comparison and magnetic-symmetry analysis reduce the library to 1347 fully compensated candidate altermagnets. **b,** Schematic illustration of the XGBoost regression model used to predict the maximum spin splitting ($\Delta E_X$) within the $E_F \pm 0.6$ eV energy window. In the gradient-boosting framework, individual decision trees are sequentially trained to correct the residual errors of the preceding ensemble, enabling progressively improved predictive accuracy. **c,** Comparison between ML-predicted and DFT-calculated $\Delta E_X$ values for the hold-out test set. The dashed line indicates ideal one-to-one agreement. The inset shows the five most important descriptors identified by SHAP feature-importance analysis, revealing the dominant roles of magnetic-site chemistry, local bonding geometry and orbital hybridization in governing the spin splitting. The model achieves strong predictive performance, with $R^2 = 0.83$, MAE = 0.17 eV and RMSE = 0.25 eV under 10-fold nested cross-validation. **d,** Dual-criterion selection map combining thermodynamic stability ($E_{hull}$) and predicted $\Delta E_X$. Grey dots represent all 1347 candidate altermagnets. Among them, 51 satisfy $E_{hull} \leq 0.1$ eV/atom, and 34 further exhibit giant spin splitting ($\geq 1.5$ eV). These compounds constitute the final high-priority candidate set for subsequent first-principles validation. **e,** Benchmarking of ML prediction against representative $AB_2C_2D$ altermagnets reported in the literature. Grey, red and green bars denote literature-extracted reference values, ML-predicted values from this work and direct DFT-calculated values, respectively. The close agreement among these values validates the transferability of the model and its reliability for identifying previously unexplored giant-spin-splitting altermagnets.

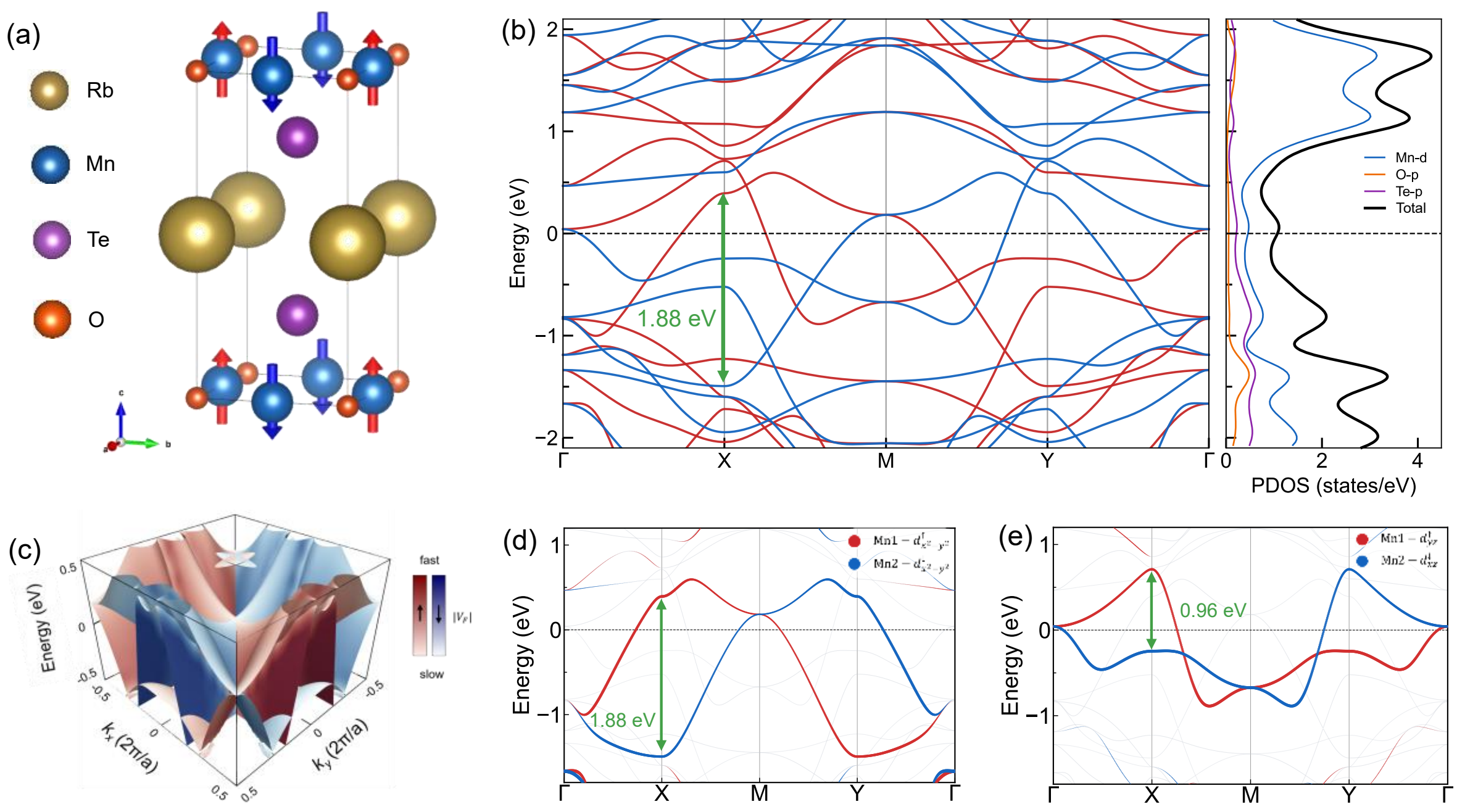

**Fig. 2 | Crystal symmetry, orbital texture and giant nonrelativistic altermagnetic spin splitting in $RbMn_2Te_2O$. a**, Fully relaxed lattice and perfectly compensated AFM ground state of $RbMn_2Te_2O$. Blue, purple, yellow and orange spheres denote Mn, Rb, Te and O atoms, respectively. Red and blue arrows indicate antiparallel local magnetic moments on the two symmetry-related Mn sublattices. **b**, Nonrelativistic spin-resolved band structure calculated without spin–orbit coupling along the Γ-X-M-Y-Γ path, together with the corresponding projected density of states (PDOS). Red and blue bands denote the two opposite spin channels, and the Fermi level is set to zero. The green arrow marks a giant altermagnetic spin splitting of approximately 1.88 eV at the X point. The PDOS shows that the states near the Fermi level are dominated by Mn-*d* orbitals, with additional ligand *p*-orbital contributions. **c**, Spin-resolved electronic structure in the $k_z = 0$ plane within an energy range from -0.5 to 0.5 eV relative to the Fermi level. The alternating red and blue regions represent opposite spin channels and reveal strongly anisotropic, d-wave-like spin texture characteristic of altermagnetism. **d,e** Sublattice- and spin-resolved fat-band projections. **d,** Projection onto the Mn1- and Mn2-derived $dx^2$-$y^2$orbitals, showing the largest spin splitting of 1.88 eV at X. **e,** Projection onto the spin-up Mn1-$d_{yz}$ and spin-dn Mn2-$d_{xz}$ orbital characters, revealing an additional spin splitting of 0.96 eV. The reciprocal locking between spin channel, Mn sublattice and orbital character directly visualizes the symmetry-enforced orbital-sublattice correspondence that drives the giant anisotropic exchange splitting.

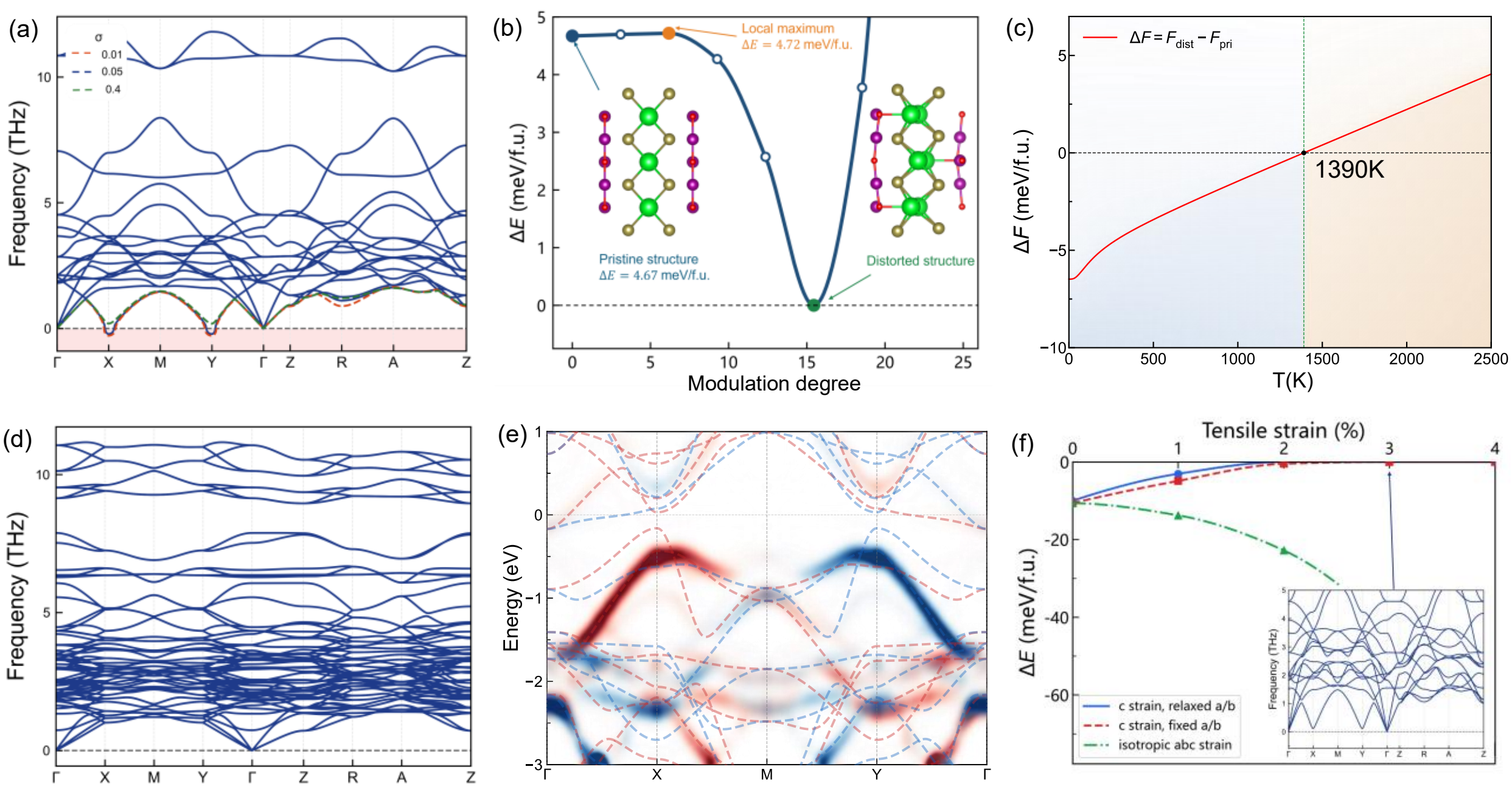

**Fig. 3 | Soft-mode-driven structural phase transition and resilient altermagnetism in $SrMn_2Te_2O$.** **a**, Phonon spectra of the pristine high-symmetry phase calculated with different electronic smearing parameters, σ = 0.01, 0.05, and 0.4 eV. Pronounced imaginary phonon modes appear near the X and Y points at small smearing, indicating a zone-boundary soft-mode instability of the 0 K pristine structure. **b**, Total-energy evolution along the X/Y soft-mode distortion coordinate. Freezing in the unstable phonon eigenvectors drives the system from the pristine high-symmetry structure into a lower-energy distorted structure, as shown by the representative structures in the insets. **c,** Temperature dependence of the Helmholtz free-energy difference, $\Delta F = F_{\mathrm{dist}} - F_{\mathrm{pri}}$, between the distorted ground state and the pristine high-symmetry phase. The crossing of $\Delta F = 0$ gives a structural transition temperature of ~1390 K, below which the distorted phase is thermodynamically favored. **d,** Phonon spectrum of the fully relaxed distorted 2 × 2 × 1 supercell. The absence of imaginary frequencies confirms that the soft-mode distortion removes the dynamical instability and stabilizes the low-temperature structure. **e,** Unfolded spin-resolved fat-band structure of the distorted 2 × 2 × 1 supercell projected onto the primitive Brillouin zone. Red and blue spectral weights denote opposite spin channels associated with the antiparallel Mn sublattices, while the original spin-resolved bands of the pristine phase are overlaid as red and blue dashed lines. The persistence of large spin-dependent spectral separation after structural reconstruction demonstrates the robustness of the altermagnetic band topology against soft-mode symmetry lowering. **f**, Tensile-strain dependence of the energy gain of the distorted phase, defined as $\Delta E = E_{\mathrm{dist}} - E_{\mathrm{pri}}$ ,under three constraints: *c*-axis strain with relaxed *a* and *b*, *c*-axis strain with fixed in-plane lattice parameters, and isotropic *abc* strain with fixed cell shape. Negative ΔE favors the distorted phase. The inset shows the phonon spectrum under 3% *c*-axis tensile strain with relaxed *a* and *b*, confirming the dynamical stability of the strained distorted structure.

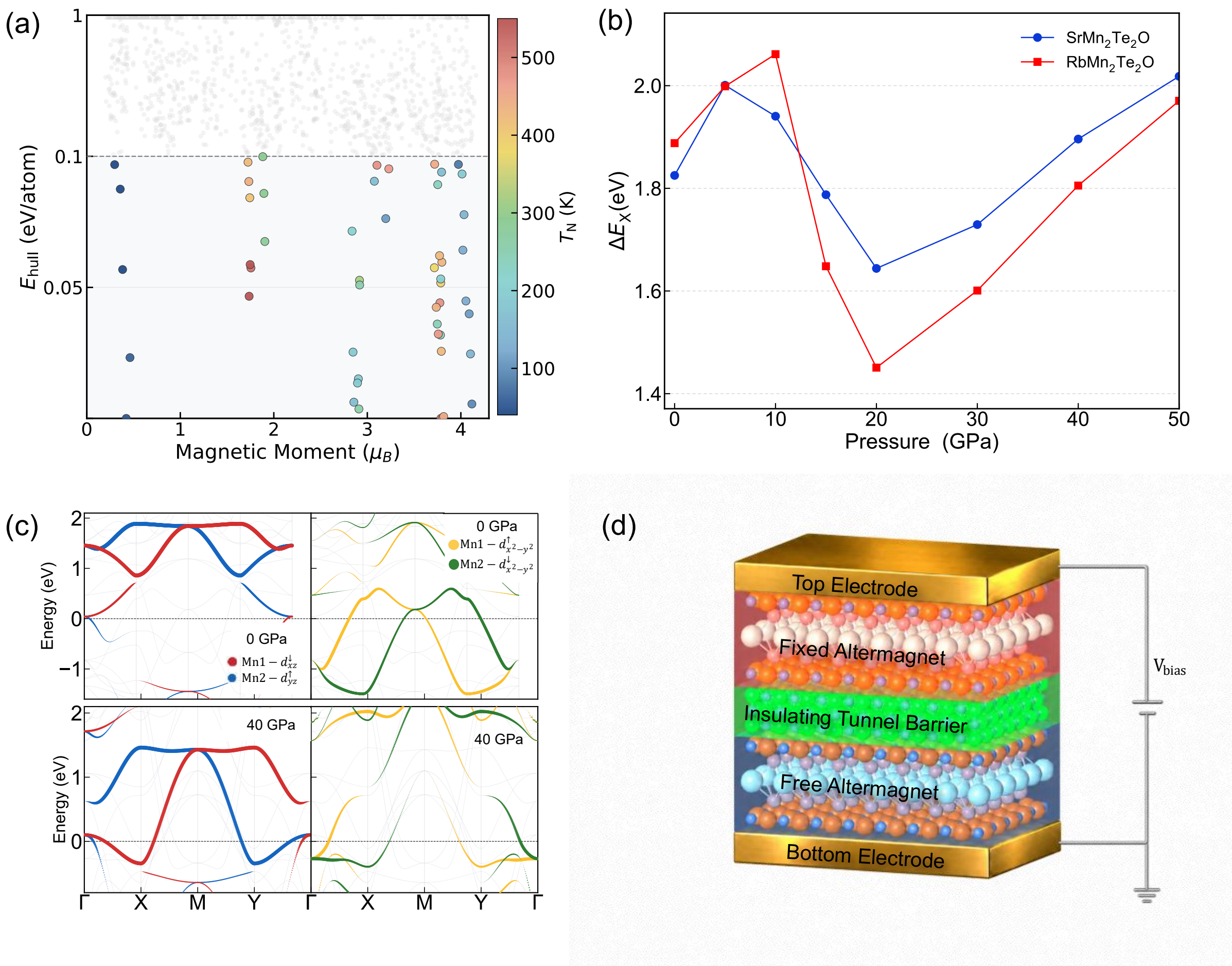

**Fig. 4 | Candidate landscape, pressure-tunable giant spin splitting and altermagnetic tunnel-junction concept. a,** Stability–magnetism map of the screened $AB_2C_2D$ altermagnetic candidates. Grey dots denote the full candidate pool, whereas colored markers highlight the highly synthesizable subset within the low-energy window ($E_{hull} \leqslant 0.1$ eV/atom). The color scale represents the estimated Néel temperature ($T_N$), revealing candidates that simultaneously combine thermodynamic stability, sizable local magnetic moments and robust finite-temperature magnetic order. **b,** Pressure dependence of the maximum spin splitting in $SrMn_2Te_2O$ and $RbMn_2Te_2O$ under hydrostatic compression. Both compounds exhibit pronounced nonmonotonic behavior, demonstrating that the giant altermagnetic splitting is not a fixed intrinsic quantity but can be actively modulated by lattice compression. **c,** Microscopic origin of the pressure tunability in $RbMn_2Te_2O$, visualized by sublattice- and spin-resolved fat-band projections at 0 and 40 GPa. The projections track the pressure-induced redistribution of Mn1- and Mn2-derived orbital characters, showing that compression substantially reshapes the band dispersion, orbital weight and sublattice-dependent exchange splitting through modified Mn-ligand hybridization. **d,** Conceptual design of an altermagnetic tunnel junction based on $AB_2C_2D$ electrodes. A generic nonmagnetic insulating tunnel barrier separates a fixed altermagnetic layer with a pinned Néel vector from a free altermagnetic layer with a switchable Néel vector. Under an applied bias $V_{bias}$, reversal of the Néel vector changes the momentum-resolved spin alignment between the two altermagnetic electrodes, thereby modulating the spin-selective tunnelling conductance. This device concept exploits the large altermagnetic spin splitting while retaining zero net magnetization, offering a route towards stray-field-free tunnelling magnetoresistance devices.

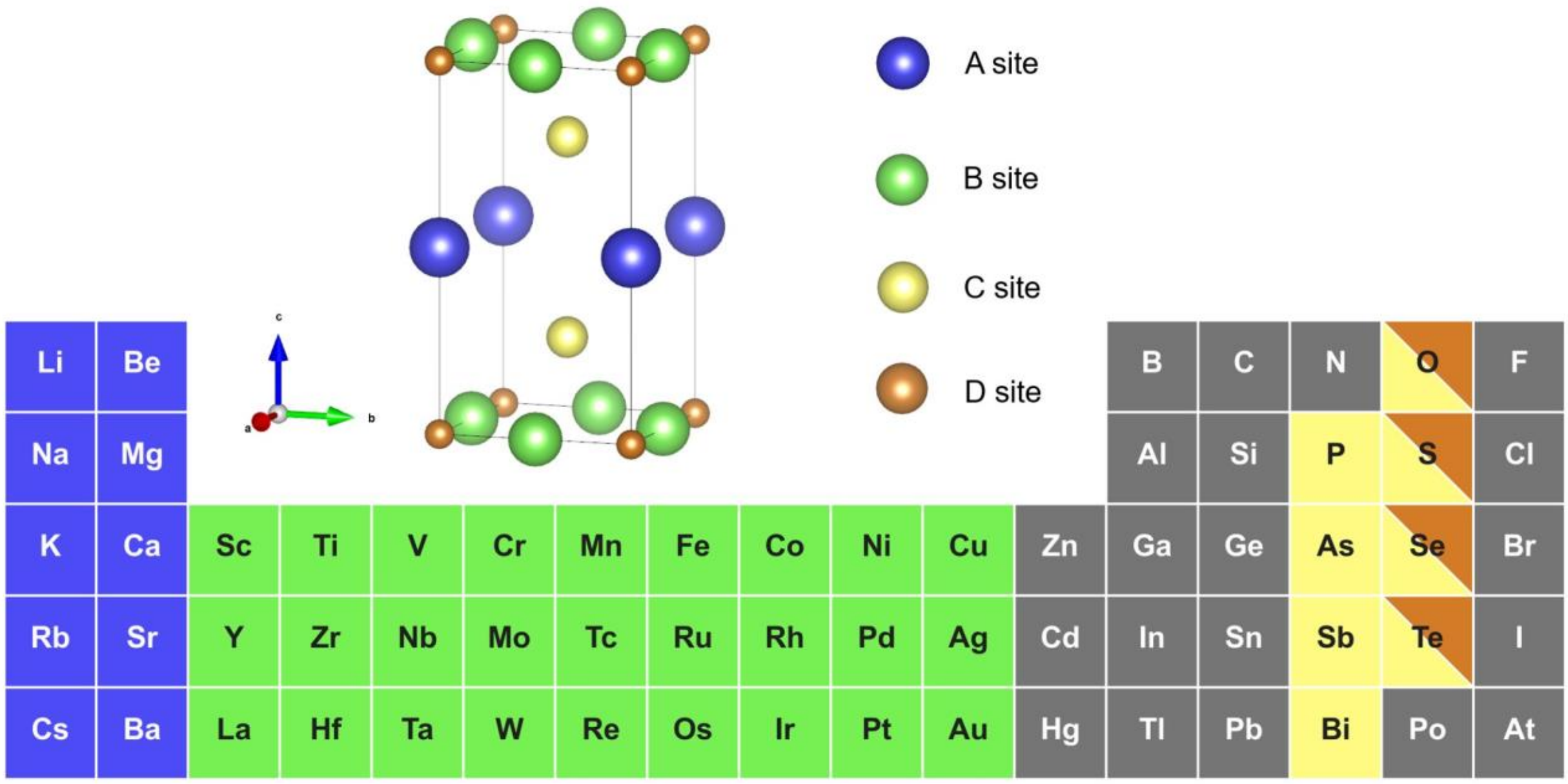

**Fig. 5 | Structural prototype and combinatorial chemical space of the $AB_2C_2D$ materials library.** The tetragonal $AB_2C_2D$ prototype structure is shown together with the site-resolved elemental substitution space used for high-throughput screening. Blue, green, yellow and orange spheres denote the crystallographically distinct A, B, C and D sites, respectively. The color-coded periodic table identifies the elements considered at each site: alkaline and alkaline-earth occupy the interlayer A site, transition metals occupy the magnetic B site, and pnictogen/chalcogen elements occupy the C and D ligand sites. The overlap between the C- and D-site substitution sets for O, S, Se and Te highlights the chemical flexibility of the anion sublattices, enabling systematic tuning of lattice geometry, local coordination, orbital hybridization and magnetic exchange across the $AB_2C_2D$ family.

| Materials | $\Delta E_X$ (eV) | $T_N$ (K) | $E_{hull}$ (eV/atom) | Materials | $\Delta E_X$ (eV) | $T_N$ (K) | $E_{hull}$ (eV/atom) |
|---|---|---|---|---|---|---|---|
| $KMn_2Se_2O^*$ | 1.71 | 490 | 0.0000 | $CsV_2Se_2O^*$ | 1.51 | 520 | 0.0574 |
| $RbMn_2Se_2O^*$ | 1.75 | 470 | 0.0000 | $NaMn_2Te_2O^*$ | 1.88 | 380 | 0.0575 |
| $RbTi_2S_2O^*$ | 1.72 | 50 | 0.0000 | $RbV_2Se_2O^*$ | 1.53 | 550 | 0.0587 |
| $CsMn_2Se_2O^*$ | 1.81 | 450 | 0.0007 | $CsMn_2Te_2O^*$ | 1.94 | 430 | 0.0596 |
| $CsCr_2Se_2O$ | 1.45 | 265 | 0.0036 | $KMn_2Te_2O^*$ | 1.85 | 425 | 0.0621 |
| $BaMn_2Se_2O^*$ | 1.62 | 95 | 0.0055 | $SrMn_2Te_2O^*$ | 1.83 | 125 | 0.0642 |
| $CsCr_2S_2O$ | 0.66 | 160 | 0.0062 | $CsV_2Te_2S^*$ | 1.76 | 280 | 0.0675 |
| $KCr_2Se_2O$ | 1.48 | 190 | 0.0135 | $KCr_2S_2O$ | 0.64 | 210 | 0.0714 |
| $RbCr_2Se_2O$ | 1.46 | 190 | 0.0151 | $BaFe_2Se_2O$ | 1.05 | 110 | 0.0755 |
| $CsTi_2S_2O^*$ | 1.70 | 60 | 0.0232 | $SrMn_2S_2O^*$ | 1.61 | 140 | 0.0767 |
| $BaMn_2S_2O^*$ | 1.61 | 140 | 0.0246 | $CsV_2Te_2O$ | 1.39 | 410 | 0.0817 |
| $RbCr_2S_2O$ | 0.66 | 180 | 0.0253 | $RbV_2Te_2S^*$ | 2.02 | 290 | 0.0830 |
| $CsMn_2S_2O^*$ | 1.78 | 425 | 0.0256 | $CsNb_2Te_2O^*$ | 1.99 | 30 | 0.0843 |
| $RbMn_2Te_2S^*$ | 1.55 | 240 | 0.0318 | $NaMn_2Te_2S^*$ | 1.54 | 230 | 0.0857 |
| $KMn_2S_2O^*$ | 1.71 | 470 | 0.0322 | $RbV_2Te_2O$ | 1.41 | 440 | 0.0867 |
| $CsMn_2Te_2S^*$ | 1.65 | 245 | 0.0360 | $CsCr_2Sb_2S$ | 1.25 | 160 | 0.0868 |
| $BaMn_2Te_2O^*$ | 1.80 | 110 | 0.0399 | $RbMn_2Te_2Se$ | 1.11 | 190 | 0.0893 |
| $NaMn_2Se_2O^*$ | 1.69 | 440 | 0.0424 | $KMn_2Se_2S$ | 1.47 | 160 | 0.0900 |
| $RbMn_2S_2O^*$ | 1.74 | 480 | 0.0441 | $BaCr_2Se_2O$ | 0.91 | 475 | 0.0913 |
| $SrMn_2Se_2O$ | 1.20 | 125 | 0.0448 | $KCr_2Te_2S$ | 1.23 | 480 | 0.0929 |
| $KV_2Se_2O^*$ | 1.56 | 550 | 0.0466 | $RbNb_2Te_2O^*$ | 2.01 | 20 | 0.0932 |
| $RbCr_2Te_2O$ | 1.37 | 255 | 0.0509 | $CsMn_2Te_2Se$ | 0.97 | 80 | 0.0933 |
| $RbMn_2Te_2O^*$ | 1.88 | 390 | 0.0517 | $NaMn_2S_2O^*$ | 1.70 | 450 | 0.0934 |
| $KCr_2Te_2O$ | 1.39 | 320 | 0.0527 | $KV_2Te_2O$ | 1.43 | 425 | 0.0945 |
| $KMn_2Te_2S^*$ | 1.51 | 200 | 0.0532 | $KV_2Te_2S^*$ | 1.81 | 290 | 0.0991 |
| $KTi_2S_2O^*$ | 1.75 | 40 | 0.0568 | | | | |

**Table S1 | Spin splitting and estimated Néel temperature of of low-energy $AB_2C_2D$ candidate materials.** Summary of the calculated altermagnetic spin-splitting magnitudes, estimated $T_N$ and $E_{hull}$ for the 51 candidate materials. Compounds marked with an asterisk denote the 34 high-priority candidates that additionally exhibit giant spin-splitting exceeding 1.5 eV.